\documentclass[prl,floatfix,amsmat,twocolumn,showpacs]{revtex4-1}
\usepackage{amssymb}
\usepackage{natbib}
\usepackage{amsmath}
\usepackage{graphicx}
\usepackage{amssymb}
\usepackage{amsmath}
\usepackage{graphicx}
\usepackage{bbm}

\def\ket #1{\vert #1\rangle}
\def\bra #1{\langle #1\vert}
\def\ketbra #1 #2 {\langle #1\vert #2 \rangle}
\DeclareMathOperator{\trace}{tr} 
\newcommand{\U}{\mathcal{U}}
\newcommand{\SU}{\mathcal{SU}}
\newcommand{\E}{\mathbb{E}}
\newcommand{\1}{\mathbbm{1}}
\newcommand{\f}{\varphi}
\renewcommand{\>}{\rangle}
\newcommand{\<}{\langle}

\begin{document}

\title{Multipartite maximally entangled states in symmetric scenarios}
\author{Carlos E. Gonz\'alez-Guill\'en}
\affiliation{Depto. de Matem\'aticas, E.T.S.I. Industriales, Universidad Polit\'ecnica de Madrid, 28006 Madrid, Spain\\
 and IMI, Universidad Complutense de Madrid, 28040 Madrid, Spain}

\begin{abstract}
We consider the class of (N+1)-partite states suitable for protocols where there is a {\it powerful} party, the authority, and the other N parties {\it play the same role}, namely the state of their system live in the symmetric Hilbert space. We show that, within this scenario, there is a ``maximally entangled state" that can be transform by a LOCC protocol into any other state. In addition, we show how to make the protocol efficiently including the construction of the state and discuss security issues for possible applications to cryptographic protocols. As an immediate consequence we recover a sequential protocol that implements the one to N symmetric cloning.
\end{abstract}

\maketitle

\section{Introduction}

The understanding, classification, quantification and use of multipartite entanglement has been one of the most challenging issues in the Theory of Quantum Information during the last decade. Even in the tripartite case, strange phenomena start to occur, like the non-equivalence of W and GHZ states \cite{DVC}, the possibility of distributing entanglement with separable states \cite{CVDC}, or the existence of unbounded violations for some correlation Bell inequalitites \cite{PWPVJ, *Briet}. Going into the N-partite situation only increases the number of interesting phenomena: universal states for quantum computation \cite{RaussendorfBriegel}, topological entanglement \cite{Kitaev}, relations with complexity theory \cite{complexity}, \ldots.

\

Associated with the different points of view in the theory of multipartite entanglement, different entanglement measures have been defined, focusing on the different aspects of entanglement: the topological entropy \cite{KitaevPreskill} measures the amount of topological entanglement in a state and is hence appropriate in the context of topological quantum computation and error correction; the localizable entanglement \cite{PVMC} measures the amount of bipartite entanglement that can be created between two sites in a collaborative scenario and is hence appropriate in the context of quantum networks and quantum repeaters; there are also measures which intend to be more general, and usually measure the distance (in some sense) to the set of separable states, like the relative entropy of entanglement, the global robustness of entanglement or the geometric measure of entanglement \cite{geometric}. As it is pointed out recursively in the literature \cite{PlenioVirmani, Horodecki}, this zoo of multipartite entanglement measures has its roots in the impossibility of defining a concept of ``maximally entangled state" in the multipartite setting.

\

 We will show here that if one imposes some symmetry restrictions to the state, motivated by the class of multipartite protocols one wants to implement with it, there is still hope to define properly the concept of a ``maximally entangled state". Here we will concentrate in protocols in which there is an authority $A$, and a set of participants $p_1,\ldots, p_N$ which {\it have to play the same role in the protocol}. This is the desired situation in a wide variety of multipartite protocols, like secret sharing or voting,  and leads to

 {\it Assumption 1:} We will work with (N+1)-partite states which are permutational-symmetric with respect to $N$ of the parties.

 {\it Assumption 2:} To make things simpler we will assume that the Hilbert space dimension of the participants is $2$, while the one of the authority will be $N+1$, which is the smallest possible dimension to purify any mixed state among the participants.

The permutational-symmetry of the state is the quantum resource, with no classical analogue, which ensures that all participants are treated equally and are indistinguishable from the authority point of view. This kind of requirements are gaining importance nowadays as {\it privacy} is really getting an issue in the new e-society. In fact, permutational symmetry also appears as a natural condition in quantum de Finetti theorems \cite{KR05}.

\

 Within assumptions 1 and 2, we will show that there is a ``maximally entangled state" $|\Phi\>$ and an LOCC protocol that transforms this to any other state with the same symmetry. Moreover, we will show how all the elements of the protocol, including the construction of the state, can be done efficiently and discuss some security issues concerning possible applications to cryptographic protocols. Along the way we will reprove the main result in \cite{cloning} from a more general point of view. We will mix basic tools from several areas: representation theory, convex analysis, Matrix Product States and quantum channels.

\section{The maximally entangled state}

\begin{figure}[t]
\includegraphics[width=4cm]{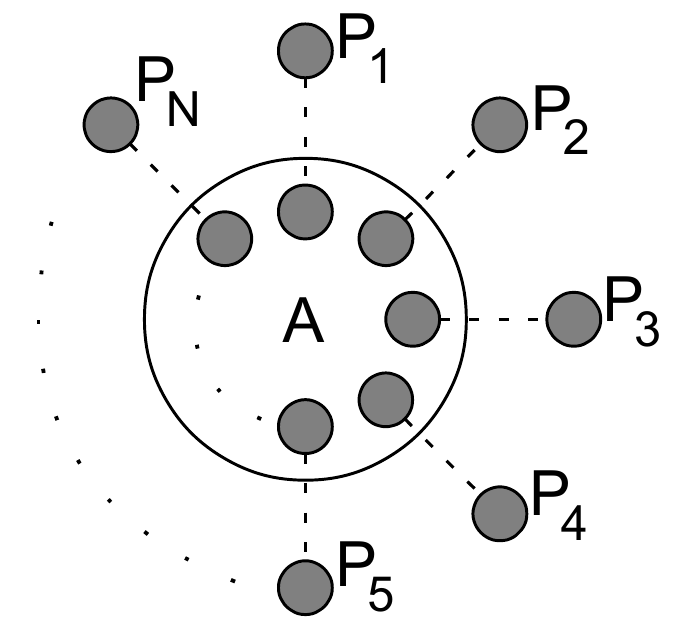}\\
\caption{Valence bond representation of the maximally entangled state $\Phi$. Solid circles connected with dotted line denote virtual EPR pairs, the big circle represents the projection in the Hilbert space ${\mathcal H}_{\rm sym}$ of the authority. \label{fig.1}}
\end{figure}

The unnormalized maximally entangled state can be described in a valence bond picture in the following way (see Fig. \ref{fig.1}). Assume that we have singlets shared between any participant and the authority. Then we project the virtual space of the authority in the permutationally symmetric subspace, which is $N+1$ dimensional. That is, we project onto the space of total spin $N/2$. This can be seen as an star-shape version of the famous AKLT state \cite{AKLT}. With a formula, our state will be:
$$|\Psi\>=(P_{\rm sym
}\otimes \mathbbm{1}_P)(|01\>-|10\>)^{\otimes N}.$$

Since we can change the singlet by any other maximally entangled state by a local unitary in any participant qubit, we can assume the same construction starting with $|00\>+|11\>$, and we will call $|\Phi\>$ to the resulting state. In most parts of this letter we will take the later. In this particular case, by considering the usual basis in the space of the authority, that is $|\alpha\> =\sum_{\substack{i_1,...,i_N\\
i_1+...+i_N=\alpha}}\frac 1 {\sqrt{\binom{N}{\alpha}}}\ket{i_1...i_N}$, we get the following explicit formula for $|\Phi\>$:
\begin{equation}\label{eq-Phi}
|\Phi\> =\sum_{\alpha=0}^N \sum_{\substack{i_1,...,i_N\\
i_1+...+i_N=\alpha}} \frac 1 {\sqrt {(N+1) \binom{N}{\alpha}}} \ket {\alpha}_A\ket{i_1,...,i_N}_P\text{.}
\end{equation}

Of course, this implies that $|\Phi\>=\frac{1}{\sqrt{N+1}}\sum_\alpha |\alpha\>_A|\alpha\>_P$ and therefore $|\Phi\>$ is the maximally entangled state along the bipartite cut AP. The problem now is that the set of participants $P$ is delocalized and therefore one cannot use general quantum operations $\E$ in $P$, but only those that are of the form $\E_1\otimes\cdots \otimes \E_N$. However, in many situations, since $V\mapsto V^{\otimes N}|_{H_{\rm sym}}$ is an irreducible representation of $\SU(2)$, Schur's lemma enables us to reduce to this situation. As we will see below, the price is the need for general POVMs, since projective measurements are no longer sufficient. In any case, the state $|\Phi\>$ is also maximally entangled in this more restrictive scenario, since one can construct from it any state with the same symmetry using only LOCC. This is the content of the following

\

{\bf Theorem 1.} There is a LOCC protocol, given below, with one way communication that allows the authority to transform $|\Phi\>$ to any known pure state $|\f\>$ that is permutationally symmetric in the Hilbert space of the participants.

\

{\bf Transformation protocol}
\begin{enumerate}
\item
Given the schmidt decomposition of the state $|\f\>=\sum_{i=0}^N\lambda_i |i\>_A |\f_i\>_P$, let $\rho^*=\sum_{i=0}^N \lambda^2_i  |\f_i^*\>\<\f_i^*|$ where $*$ means complex conjugation. The authority $A$ measures with measurement operators $\{F_U=\sqrt{N+1}\pi(U) \rho^* \pi(U^{\dag})\}$ his part of the system where the $U$s are
distributed with respect to the Haar measure in $\mathcal{SU}(2)$ and $\pi$ is the (unique) irreducible representation of $\SU(2)$ in a $N+1$ dimensional space given by $V\mapsto V^{\otimes N}|_{\mathcal{H}_{\rm sym}}$. 
\item
$A$ broadcasts the result of the measure $U_0$.
\item
Each participant applies to his system the unitary $YU_{0}Y$ to obtain the state $|\f\>$.
\end{enumerate}

This Theorem shows also that our state could be of use in situations (like secret sharing or key distribution) in which one authority is assumed to distribute some quantum state among the set of participants. One advantage now is that only permutationally symmetric states can be constructed and all the participants are then sure that they are treated in equal footing.

\

{\it Proof of the Theorem.}

The result relies essentially on Schur's lemma, which guarantees that the measure in step 1 of the protocol is indeed a measure since \begin{equation}\label{1}
\frac{1}{N+1}\mathbbm{1}_{H_{A}}=\int_{\mathcal{U}(2)} \pi(U) \rho^*
\pi(U^\dag) dU\; .
\end{equation}

It only remains to show that the state after the protocol is the one
we want, which is a routine calculation. Suppose the result of the measure is $\alpha$, then the state after the measure reads \begin{equation}\label{output-state-1}\left( \pi(U_\alpha)\sum_{i=0}^N \lambda_i  |\f_i^*\>\<\f_i^*|  \pi(U^\dag_\alpha) \otimes \mathbbm{1}_{P} \right)|\Phi\>.\end{equation} Now, by the definition of $\pi$ and the fact that $|00\>+|11\>$ is $U \otimes YUY$ invariant for any $U\in \U(2)$, we get that $ \pi(U) \otimes (YUY)^{\otimes N}|\Phi\>=|\Phi\>$ for every $U$. Using  (\ref{eq-Phi}) is now trivial to conclude that (\ref{output-state-1}) is indeed equal to  \begin{equation}\label{output-state-2}(\pi(U_\alpha)\otimes (YU_\alpha Y)^{\otimes N}) \sum_{i=0}^N \lambda_i |\f_i^*\>_A |\f_i\>_P.\end{equation}
Therefore, after knowing the result $\alpha$, each participant can apply $Y U_\alpha^\dag Y$ to his system and to obtain the joint state $|\f\>$ and $A$ can apply the unitary that takes $\pi(U_\alpha) |\f_i^*\>$ to $|i\>$ $\square$.

\

Considering $|\f\>$ to be a product state between the authority and the participants we have

\

{\bf Corollary 2 (State-transfer).} Given $|\Phi\>$, there is a LOCC protocol, given below,  with one way communication that allows the authority to create in the Hilbert space of the participants any permutationally symmetric pure state $|\f\>$.

\

The first thing to notice here is that the measurement required in step 1 of the state-transfer protocol has an infinity number of outcomes, which in turns implies that one needs an infinite dimensional ancilla in order to implement it with orthogonal projectors. The way around this problem is by considering a set of unitaries $\{U_i\}_{i=1}^k \subset \U(2)$ and a set of scalars $\omega_i\ge 0$ such that $\sum_{i}\omega_i=1$ and
\begin{equation}\label{design} \sum_{i=1}^k\omega_i \pi(U_i)\rho^*\pi(U_i^{\dag})=\int_{\SU(2)}\pi(U) \rho^* \pi(U^{\dag}).
\end{equation} This allows to replace the measurement in step 1 of the protocol by the one with operators $\{F_i=\sqrt{\omega_i(N+1)} \pi(U_i)\sum_{j=0}^N \lambda_j \ket{ \f_j^*}\bra{\f_j^*}\pi(U_i^\dag)\}$.
Using Caratheodory's Theorem it is not difficult to show that, in this case, $k$ can indeed be taken $\le (N+1)^2+1$ and hence polynomial in $N$ (see Appendix).

Since in step 2, the authority will broadcast the outcome of the measurement, it is interesting to note that, from (\ref{output-state-2}), the probability of obtaining the output $i$ is $\omega_i$ and hence independent of the state $|\f\>$ being transferred. This is crucial in cryptographic applications, like secret sharing, in which the public communication should give no information at all. The main problem with this state-transfer protocol is that the measurement in A, although being local, depends on the state to transfer, and therefore it does not work in situations in which the authority wants to transfer an unknown state. However, thanks to Schur's lemma, it is possible to design a teleportation-like protocol that also works under our assumptions and allows A to teleport with LOCC any permutationally symmetric unknown state to P. The procedure is a particular case of the situation described in \cite{teleportation-general} and can be resumed in:

\

{\bf Teleportation-like protocol}
\begin{enumerate}
\item The initial joint system is $|\f\>_{A_1}\otimes |\Phi\>_{A_2P}$, where $|\f\>$ is the state to be teleported.
\item
The authority $A$ measures with measurement operators
$\{F_U=(N+1)\pi(U)_{A_1}\otimes \1_{A_2} |\Phi\>\<\Phi| \pi(U^{\dag})_{A_1}\otimes \1_{A_2}\}$ his part of the system where the $U$s are
distributed with respect to the Haar measure in $\mathcal{U}(2)$
\item
$A$ broadcasts the result of the measure $U_0$.
\item
Each participant applies to his system the unitary $U_{0}$ to obtain the state $|\f\>$.
\end{enumerate}

\

Exactly as before, one can use a discrete set of unitaries to avoid the continuous parameter. In this case equation (\ref{design}) should hold for any matrix $\rho \in \mathcal{M}_{N+1}$. By a similar reasoning one can show that the number of unitaries needed is upper bounded by $4(N+1)^4+1$. Nevertheless, weighted N-designs in $\U(2)$ already solve this problem and such a design exists with $\le \left(
\begin{array}{c}
2N+3   \\
3   
\end{array}
\right)$
unitaries \cite{RoyScott}. Likely, not only the output of the measurement is completely independent of the state to be teleported, but also the set of unitaries itself. Finally, it is trivial to see that the same protocol allows to teleport arbitrary unknown mixed states supported on the symmetric subspace.

At the light of this result, it seems that if we restrict to our Assumptions 1 and 2 everything works essentially as in the bipartite case, in which we start with the maximally entangled state $|\Phi\>=\sum_\alpha |\alpha\>_A|\alpha\>_P$. As we commented above, there is at least one  important difference. In the protocols presented here we use POVMs instead of projective measurements. It is interesting to note that it is indeed {\it  impossible} to reduce to projective measurements, as it is shown in the following

\

{\bf Theorem 3.} It is not possible to implement the teleportation-like protocol using projective measurements.

\

{\it Proof:}
Let us assume that it is possible to teleport from A to P the unknown permutationally symmetric state $|\f \>$ with projective measurements. It implies that there must exist a decomposition of the form
$$|\f\>_{A_1}|\Phi\>_{A_2P}=\sum_{r\in R} \sqrt{p_r}|r\>_{A}\otimes \pi(U_r)|\f\>,$$
where $|r\>$ is an orthonormal set in the joint system $A=A_1A_2$.  On one hand, if we trace out system $A$ we get $\1_{H_{\rm sym}}=\sum_r \pi(U_r)|\f\>\<\f|\pi(U_r)^{\dagger}$, which implies that $|R|\ge N+1$.  On the other hand, if we trace out system P, we get
$|\f\>\<\f|\otimes \1_{\mathcal{H}_{\rm sym}}=\sum_{r,s\in R} \<\f|\pi(U_s^{\dag}U_r)|\f\> |r\>\<s|, $ which implies that
$$\1_{\mathcal{H}_{\rm sym}\otimes \mathcal{H}_{\rm sym}}= \sum_{r,s} \trace(U_s^\dag U_r)^N |r\>\<s|$$
and hence $\trace(U_s^\dag U_r)=\delta_{rs}$. But this is not possible since $U_r,U_s\in \U(2)$ \; $\square$.

\section{Properties of the state}

\subsection{Characterization by symmetries}
Just as the state $|00\>+|11\>$ can be characterized as the unique pure two-qubit state that is invariant under the action of $U\otimes U$ for any unitary $U$, one can show that our state $|\Phi\>$ is the unique pure state, within Assumptions 1 and 2, that is invariant under the action of $U^{\otimes N} \otimes \pi(YUY)$ for any unitary $U\in \U(2)$, where $\pi$ is the (unique) unitary irreducible representation of $\SU(2)$ in an $N+1$ dimensional space given by $V\mapsto V^{\otimes N}|_{\mathcal{H}_{\rm sym}}$.

\begin{figure}[t]
\includegraphics[width=8cm]{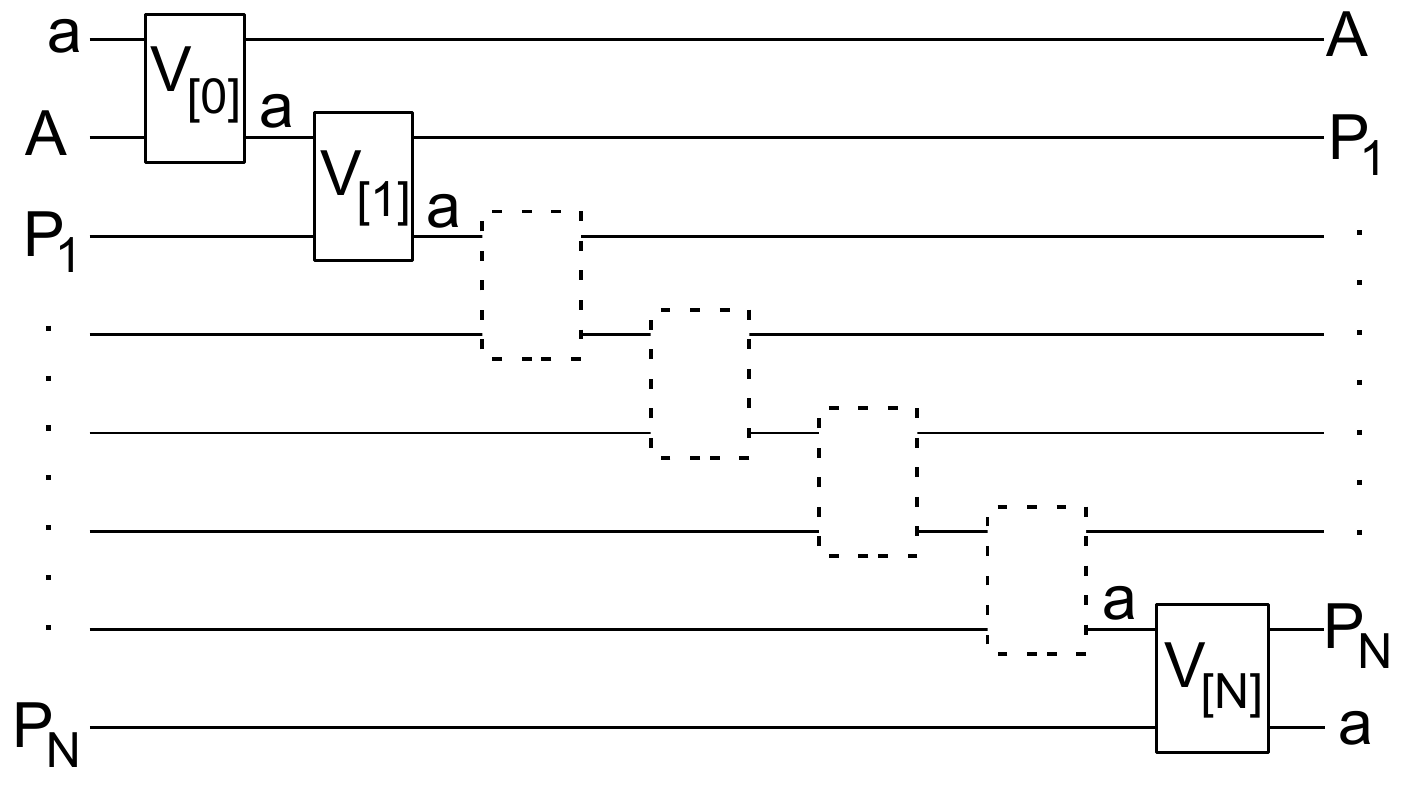}\\
\caption{Circuit for creating the state $\Phi$ in a sequential way, where each box implements a unitary $V_{[i]}$ between the ancilla $a$ and participant $i$ together with a swap operation between them. \label{fig2}}
\end{figure}

\subsection{Creation of the state}
Is there an efficient way, that is, polynomial in the parameters, to construct the 'maximally entangled state' $|\Phi\>$? The answer is yes and comes from the following Matrix Product State representation:
$$|\Phi\>=\sum_{\alpha_0,i_1,\ldots,i_N}A^{[0]}_{\alpha_0}A^{[1]}_{i_1}\cdots A^{[N]}_{i_N}|\alpha_0i_1\cdots i_N\>,$$
where 
\begin{align*}
A_{\alpha_0}^{[0]}\hspace{-0,1cm}&=(0,...,0,\underset{\alpha_0}{1},0,...,0),\\
A_{i_j}^{[j]}\hspace{-0,1cm}&=\hspace{-0,2cm} \sum_{\alpha_j=0}^{N-j} \hspace{-0,1cm} \frac
{\sqrt{(N-j+1)\binom{N-j}{\alpha_j}}}
        {\sqrt  {(N-j+2)\binom{N-j+1}{\alpha_j+i_j}}}  \ket{\alpha_j+i_j}\bra{\alpha_j}, \; j\hspace{-0,1cm}=\hspace{-0,1cm}1,...,N\;\hspace{-0,1cm}.\end{align*}
Using the result in \cite{PVWC06}, this immediately gives an efficient way to create the state $|\Phi\>$ in the following sequential manner (see Fig. \ref{fig2}):
$$|\Phi\>_{AP}|0\>_a=V_{[0]}\cdots V_{[N]}|0\cdots 0\>_{AP} |0\>_a,$$
where $a$ is an ancillary system of dimension N+1 and $V_{[j]}$ is the unitary gate, involving only participant $j$ ($0$ being the authority) and the ancilla $a$, given by $$V_{[j]}|0\>_{j}|\alpha\>_a=\sum_{i_j}\<s|A^{[j]}_{i_j}|r\> |i_j\>_j|s\>_a.$$ The condition $\sum_{i_j}A^{[j]\dag}_{i_j}A^{[j]}_{i_j}=\1$ makes $V_{[j]}$ unitary \cite{PVWC06}.
Of course, one may take the authority system as the ancilla and then obtain the state $|\Phi\>$ after one round of two body interactions between the authority and each participant.

\subsection{Sequential cloning}

\begin{figure}[t]
\includegraphics[width=8cm]{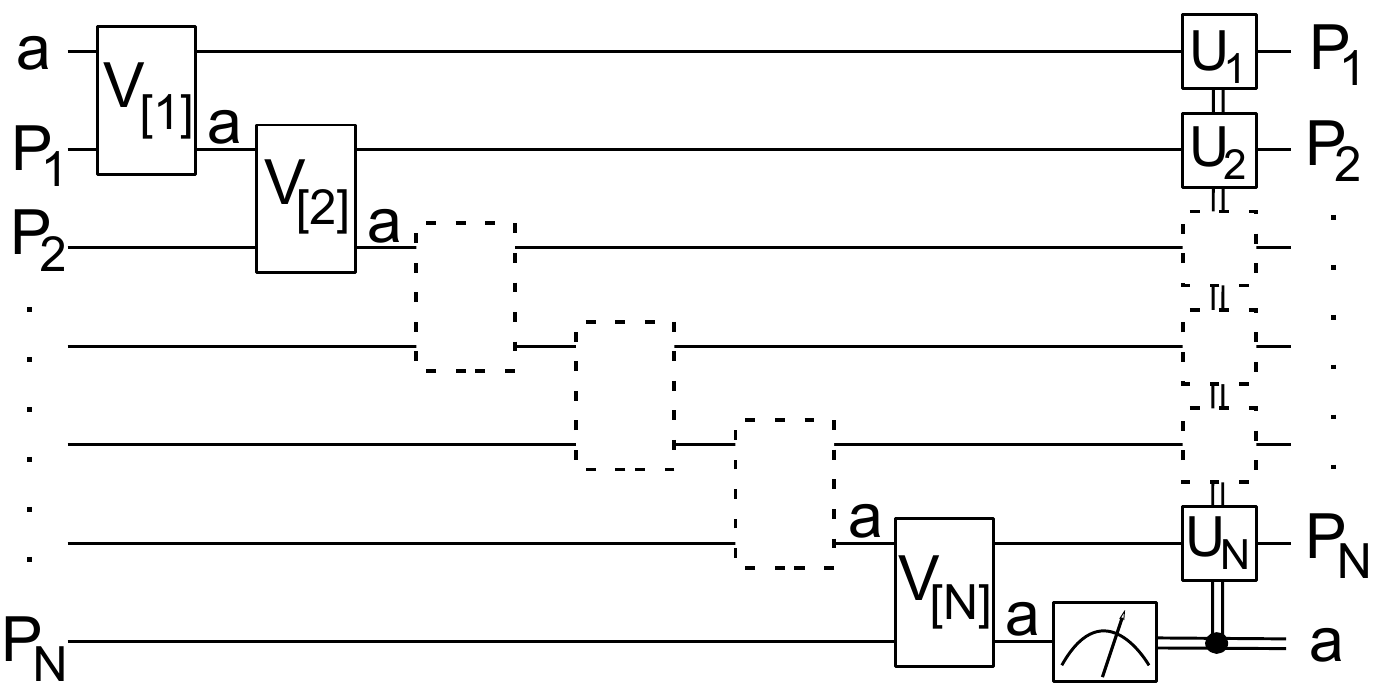}\\
\caption{Sequential circuit implementing cloning. Firstly, the maximally entangled state is created sequentially between the ancilla and the participants. Secondly, the ancilla is measured. Finally, depending on the result of the measurement, local unitary corrections are applied in the participants.\label{fig3}}
\end{figure}

The fundamental no-cloning theorem  \cite{no-cloning} states that it is impossible to clone unkonwn quantum states. However, as one can infer from the excellent review \cite{review-cloning}, there are many situations in cryptography in which the optimal approximate cloning is important. In \cite{cloning} (see \cite{sequential} for a refinement), the authors use Matrix Product State theory to design a protocol which implements the $1\rightarrow N$ symmetric universal quantum cloning in the following sequential manner (see Fig. \ref{fig3}).

\begin{enumerate}
\item[Step 1] An ancilla of dimension $O(N)$ interacts sequentially with each qubit.
\item[Step 2] A final measurement is implemented in the ancilla.
\item[Step 3] A local unitary correction is made in the qubits depending on the output of the measurement.
\end{enumerate}

Since in the symmetric universal cloning the final state is supported in the symmetric subspace one can use step 1 to create our maximally entangled state and steps 2,3 to teleport the cloned state to all the qubits with our teleportation-like protocol. Of course, the same can be done for {\it any} protocol in which the final state lives in the symmetric subspace.

\section{Checking symmetry}
Since the participants want to keep their privacy, they must have a way to be sure that the state they receive from the authority is permutational-symmetric or, even more, is supported in the symmetric subspace. The latter is indeed equivalent to implement the measure of the total spin in $N$ spin-$\frac{1}{2}$ particles. A simple way to do so is the following protocol which requires very few computational power to the participants: 1 qubit channel from participant i to participant i+1 and the ability of implementing 2-qubit measures.
The protocol aims to (i) do nothing if the original state was supported on the symmetric subspace, (ii) end up with a state supported on the symmetric subspace.

The protocol repeats $R$ times the following round. With probability $1/N$ participant $i$ send his qubit to participant $i+1$ which check if the $i, i+1$ qubits are supported in the symmetric or the antisymmetric subspace. If it is the latter, he constructs the mixed state over the symmetric subspace and sends the $i$-th qubit back to participant $i$. 

The quantum channel implemented is $\Phi=1/N\sum_{i=1}^N  T_{i,i+1} \otimes \1_{All \backslash i,i+1} $ where $T_{i,i+1}(\rho)=P_{\rm sym}\rho_{i,i+1} P_{\rm sym} + 1/4\<\Psi_{-}|\rho_{i,i+1}|\Psi_{-}\> \1$. It is clear that this channel verifies (i). (ii) is consequence of the fact that all fixed points of $\Phi$ are supported in the symmetric subspace. To see this we rely on \cite{QIEP}, which characterizes the fixed points as those matrices $\rho$ that are fixed point of $T_{i,i+1}$ for any $i=1,...,N$, these are density matrices that are supported in the symmetric subspace of any pair of consecutive participants.

\begin{figure}[t]
\includegraphics[width=8.5cm]{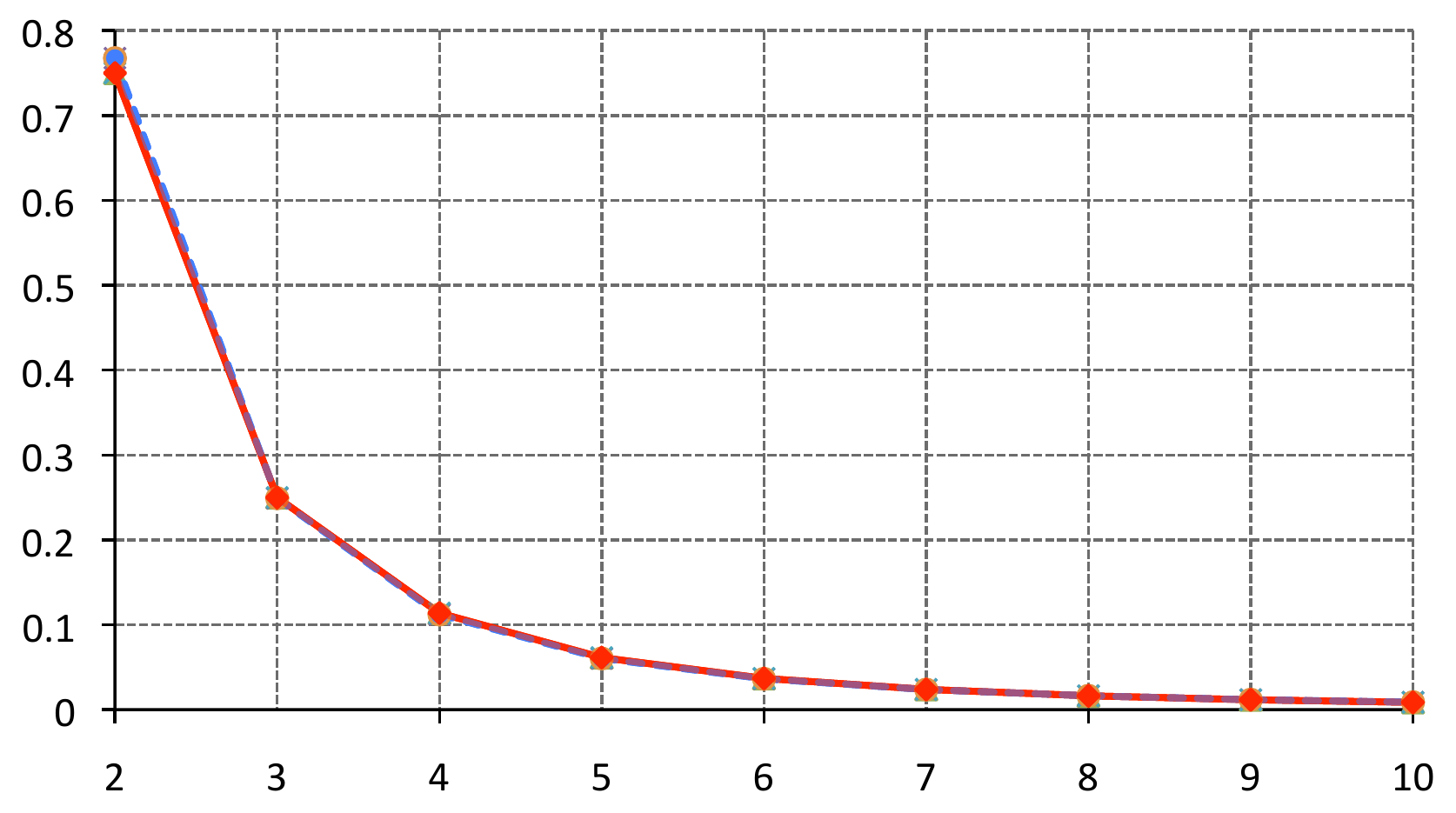}\\
\caption{Gap of $\Phi$ as a function on the number of participants (red) compared with the values of its regression function when considering a power regression model (blue). The exponent of the function is $-2.77$. \label{fig4}}
\end{figure}

The efficiency of the protocol, that is, how it approaches a fixed point with the number of iterations is governed by the modulus of the second largest eigenvalue of $\Phi$. Numerically (see Fig. \ref{fig4}) the second eigenvalue of the protocol after $O(N^3)$ rounds seems to behave as $(1-cN^{-2.77})^{O(N^3)}$, where $c$ is a constant, which is exponentially small in $N$.

Alternatively one can use the general procedures concerning secure multipartite quantum computation in \cite{multiparty,*multiparty2}.

\section{Conclusion}

We have considered the set of multipartite states in which the system of the participants live in their symmetric subspace and whose state is purified by the authority. Among this set we find a maximally entangled state which, thanks to Schur lemma, can be transformed into any other and allows to make teleportation from the authority to the participants. Nevertheless, POVMs are needed for both applications. We have shown how to create this maximally entangled state sequentially in an efficient way thanks to its Matrix Product State representation. Putting together the sequential generation and the teleportation result, we reprove that any protocol in which the final state lives in the symmetric subspace can be done sequentially in an efficient way. This is illustrated with the $1\rightarrow N$ symmetric universal quantum cloning. Moreover, we have argued that the result of the measures in the protocols does not reveal information and that the participants can make sure that their state lives in the symmetric subspace.

\section{Acknowledgements}

We thank Sofyan Iblisdir, Juanjo Garc\'ia-Ripoll and David P\'erez-Garc\'ia for their useful comments and discussions. This work has been partially funded by the Spanish grants MTM2011-26912 and QUITEMAD and the European project QUEVADIS. 

\bibliography{bibMMES}

\appendix

\section{Appendix}

We show in this appendix that given a density matrix $\rho \in \text{Herm}({\mathcal H}_{\rm sym})$ there exists a set of unitaries $\{U_i\}_{i=0}^{(N+1)^2} \subset \U(2)$ and a set of scalars $\lambda_i\ge 0$ such that $\sum_{i}\lambda_i=1$ and 
$$\sum_{i=0}^{(N+1)^2}\lambda_i \pi(U_i) \rho \pi(U_i^{\dag})=\int_{\SU(2)}\pi(U) \rho \pi(U^{\dag})$$
{\it Proof:} Let $T:\mathbb{U}(2) \rightarrow \text{Herm}({\mathcal H}_{\rm sym})$ be
defined by $U\rightarrow T_U$ where $T_U=U^{\otimes N} \rho U^{\otimes N
\dag}$. Let $S \in \text{Herm}({\mathcal H}_{\rm sym})$ be the result of applying the twirling operator to $\rho$, that is,
$S:=\int_{U\in\mathcal{U}(2)} U^{\otimes N} \rho U^{\otimes N
\dag}dU$. We have shown in (\ref{1}) that $S= \frac{1}{N+1}\mathbb{I}_{{\mathcal H}_{\rm sym}}.$ Let h be a linear functional of $ \text{Herm}({\mathcal H}_{\rm sym})$ whose positive closed half-space contains $\text{Im}(T)$ then 
$$h(S)=\int_{U\in\mathcal{U}(2)} h(T_U)dU\geq 0,$$
so $h(S) \in  \overline{co}(\text{Im}(T))$. Moreover, if $Im(T)\nsubseteq ker(h)$ then there is an $\epsilon>0$ such that $\text{Im}(T)$ meets $h^{-1}((\epsilon,\infty))$. So the set $V=(h\circ T)^{-1}((\epsilon,\infty)) $ of $\mathbb{U}(2)$ is nonvoid and, by the continuity of T, open. Therefore $h(S)>\epsilon \mu (V)>0.$ Hence 
$$S\in \text{relint} co(\text{Im} (T))\subseteq co(\text{Im}(T)).$$
Then, applying Caratheodory's theorem \cite{Cara}, there exist functions $U_0,U_1,...,U_{(N+1)^2} \in \mathcal{U}(2)$ such that $S\in co\{T_{U_0},T_{U_1},...,T_{U_{(N+1)^2}}\}.$ That is, there exist $\lambda_i\geq 0$, $U_i\in\mathcal{U}(2)$ for $i=0,...,(N+1)^2$ such that $\sum_{i}\lambda_i=1$ and 
$$\frac{tr(A)}{N+1}\mathbb{I}_{{\mathcal H}_{\rm sym}}=\sum_{i=0}^{(N+1)^2} \lambda_i U_i^{\otimes N} \rho U_i^{\otimes N \dag}$$
$\square$.

\end{document}